\documentstyle{ioplppt}
\begin{document}
\def\T{{\cal T}}
\def\D{{\cal D}}
\jl{1}

\letter{$N$-Soliton Solutions to a New ($2 + 1$) Dimensional
Integrable Equation}

\author{YU Song-Ju\ftnote{1}{fpc30017@bkc.ritsumei.ac.jp}, Kouichi
TODA\ftnote{2}{sph20063@bkc.ritsumei.ac.jp} and Takeshi
FUKUYAMA\ftnote{4}{fukuyama@bkc.ritsumei.ac.jp}}

\address{Department of Physics, Ritsumeikan University, Kusatsu, Shiga
525 JAPAN}

\begin{abstract}
We give explicitly $N$-soliton solutions of a new ($2 + 1$)
dimensional equation, $\phi_{xt} + \phi_{xxxz}/4 + \phi_x \phi_{xz} +
\phi_{xx} \phi_z/2 + \partial_x^{-1} \phi_{zzz}/4 = 0$.  This equation is 
obtained by unifying two directional generalization of the KdV equation, 
composing the closed ring with the KP equation and Bogoyavlenskii-Schiff 
equation.  We also find the Miura transformation which yields the same 
ring in the corresponding modified equations.
\end{abstract}

\maketitle

The study of higher dimensional integrable system is one of the central 
themes in integrable systems.  A typical example of higher dimensional 
integrable systems is to modify the Lax operators of a basic equation, in 

this letter the potential KdV(p-KdV) equation.  The Lax pair of the p-KdV 
equation have the form 

\begin{eqnarray}
& &L(x,t) = \partial_x^2 + \phi_x(x,t), \label{Lofkdv} \\ 
& &T(x,t) = \biggl(L(x,t)^{\frac{3}{2}}\biggr)_+ + \partial_t = 
\partial_x L(x,t) + T'(x,t) + \partial_t.
\label{Tofkdv}
\end{eqnarray}
The p-KdV equation

\begin{equation}
\phi_{xt} + \frac{1}{4} \phi_{xxxx} + \frac{3}{2} \phi_x \phi_{xx} = 0,
\label{kdveq}
\end{equation}
is equivalent to the Lax equation

\begin{equation}
[L, T] = 0.
\label{laxeq}
\end{equation}
B. G. Konopelchenko and V. G. Dubrovsky modified $L$ operator from 
($1 + 1$)-dimensions to ($2 + 1$)-dimensions and gave the Lax pair 
of the KP equation\cite{kd}.  On the other hand, O. I. Bogoyavlenskii 
modified $T$ operator and gave the Lax pair of the 
Bogoyavlenskii-Schiff(BS) equation\cite{b}.  We modified $L$ and $T$ 
operators searching for ($3 + 1$)-dimensional integrable equation.  
However, the Lax equation was eventually reduced to ($2 + 1$)-dimensional 
equation\cite{fty1},

\begin{equation}
\phi_{xt} + \frac{1}{4} \phi_{xxxz} + \phi_x \phi_{xz} + \frac{1}{2}
\phi_{xx} \phi_z + \frac{1}{4} \partial_x^{-1} \phi_{zzz} = 0.
\label{neweq}
\end{equation}

In this letter we will give explicitly $N$-soliton solutions to this 
new ($2 + 1$)-dimensional equation.  Moreover, we will give the modified 
equation of (\ref{neweq}) from the Miura transformation and its Lax pair.  

Equation(\ref{neweq}) admits the Lax 
representation(\ref{laxeq})\cite{b,fty1}:

\begin{eqnarray}
& &L = \partial_x^2 + \phi_x + \partial_z, \label{L} \\ & &T =
\partial_x^2 \partial_z + \frac{1}{2} \partial_z^2 + \frac{1}{2}
\phi_z \partial_x + \phi_x \partial_z + \frac{3}{4} \phi_{xz} -
\frac{1}{4} \partial_x^{-1} \phi_{xz} + \partial_t. \label{T}
\end{eqnarray}
Equation(\ref{neweq}) has also the Painlev\'e property\cite{fty1} in 
the sense of WTC method\cite{wtc}.

By the dependent variable transformation

\begin{equation}
\phi \equiv 2 \frac{\tau_x}{\tau},
\label{depvartf}
\end{equation}
equation(\ref{neweq}) is transformed into the trilinear form

\begin{equation}
(36 \T_x^2 \T_t + \T_x^4 \T_z^{\ast} + 8 \T_x^3 \T_x^{\ast} \T_z + 9
\T_z^3) \tau \cdot \tau \cdot \tau = 0.
\label{trilinearneweq}
\end{equation}
The operators $\T$, $\T^{\ast}$ are defined by\cite{grh,hgr}

\begin{equation}
\fl \T_z^n f(z) \cdot g(z) \cdot h(z) \equiv (\partial_{z_1} + j
\partial_{z_2} + j^2
\partial_{z_3})^nf(z_1)g(z_2)h(z_3)|_{z_1=z_2=z_3=z},
\label{Tdef}
\end{equation}
where $j$ is the cubic root of unity, $j = \exp(2 \i\pi/3)$.  
$\T^{\ast}_z$ is the complex conjugate operator of $\T_z$ obtained by 
replacing $(\partial_{z_1} + j \partial_{z_2} + j^2\partial_{z_3})$ by 
$(\partial_{z_1} + j^2 \partial_{z_2} + j\partial_{z_3})$.  
Equation(\ref{trilinearneweq}) was obtained by J. Hietarinta, 
B. Grammaticos and A. Ramani from the singularity analysis of the 
trilinear equation\cite{hgr}.  The $1$-soliton solution of 
equation(\ref{trilinearneweq}) takes the form

\begin{equation}
\tau_1 = 1 + \exp(P_1 x + Q_1 z + R_1 t + S_1),
\label{1ss}
\end{equation}
and the dispersion relation is

\begin{equation}
P_1^4 Q_1 + Q_1^3 + 4 P_1^2 R_1 = 0.
\label{disprel}
\end{equation}
Here $S_1$ is a constant.  Nextly the form of $2$-soliton solution is
written as

\begin{eqnarray}
\fl \tau_2 &=& 1 + \exp(P_1 x + Q_1 z + R_1 t + S_1) + \exp(P_2 x +
Q_2 z + R_2 t + S_2) \nonumber \\ \fl &+& A_{12} \exp\bigl((P_1 + P_2)
x + (Q_1 + Q_2) z + (R_1 + R_2) t + S_1 + S_2\bigr), \label{2ss}
\end{eqnarray}
and $A_{12}$ is

\begin{equation}
A_{12} = \frac{P_1^2 P_2^2 (P_1 - P_2)^2 - (P_1 Q_2 - P_2 Q_1)^2}
{P_1^2 P_2^2 (P_1 + P_2)^2 - (P_1 Q_2 - P_2 Q_1)^2}.
\label{zure}
\end{equation}
As a result, we have the conjecture that $N$-soliton solutions of
equation(\ref{trilinearneweq}) have the form 

\begin{eqnarray}
& &\tau_N = 1 + \sum^N_{n = 1} \sum_{{}_NC_n} A_{i_1 \cdots i_n}
\exp(\eta_{i_1} + \cdots + \eta_{i_n}), \label{nssofneweq} \\ 
& &\eta_j = P_j x + Q_j z + R_j t + S_j, 
\hspace{0.5cm} P_j^4 Q_j + Q_j^3 + 4 P_j^2 R_j = 0 \label{etaneweq} \\ 
& &A_{jk} = \frac{P_j^2 P_k^2 (P_j - P_k)^2 - (P_j Q_k - P_k Q_j)^2} 
{P_j^2 P_k^2 (P_j + P_k)^2 - (P_j Q_k - P_k Q_j)^2}, \label{Ajk} \\ 
& &A_{i_1 \cdots i_n} = A_{i_1,i_2} \cdots A_{i_1,i_n} \cdots 
A_{i_{n - 1},i_n}. \label{Aijklm}
\end{eqnarray}
Here the summation ${}_NC_n$ indicates summation over all possible 
combinations of $n$ elements taken from $N$, and symbols $S_j$ always 
denote arbitrary constants.  However, this equation also allows the 
following the Wronskian type solution(\ref{wronsolneweq}), which is 
easier for the analytic proof of solution than the form(\ref{nssofneweq}).  
We introduce new parameters $p_j$ and $q_j$ as follows,

\begin{equation}
p_j - q_j = P_j, \hspace{0.5cm} p_j^2 - q_j^2 = \pm Q_j,
\hspace{0.5cm} p_j^4 - q_j^4 = \mp 2 R_j. \label{newparas}
\end{equation}
We rewrite $N$-soliton solution $\tau_N$(\ref{nssofneweq}) in the form of 
$N \times N$ Wronskian

\begin{equation}
\tau_N = \det \pmatrix{ f_1    & \cdots & \partial^{N - 1}_x f_1 \cr
                        \vdots & \cdots & \vdots                 \cr 
                        f_N    & \cdots & \partial^{N - 1}_x f_N \cr }, 
\label{wronsolneweq}
\end{equation}
where

\begin{equation}
f_j = \exp\bigl(p_j x \pm p_j^2 z \mp \frac{1}{2} p_j^4 t + c_j\bigr) + 
\exp\bigl(q_j x \pm q_j^2 z \mp \frac{1}{2} q_j^4 t + d_j\bigr)
\label{elements}
\end{equation}
with constants, $c_j$ and $d_j$.  We prove analytically that Wronskian 
form(\ref{wronsolneweq}) is indeed the solution to a new 
($2 + 1$)-dimensional trilinear equation(\ref{trilinearneweq}).  For later 
use, we represent the Wronskian solution(\ref{wronsolneweq}) as follows, 

\begin{equation}
\tau_N = \det \pmatrix{
  f_1    & \cdots  & \partial^{N - 1}_x f_1 \cr
  \vdots & \cdots  & \vdots                 \cr
  f_N    & \cdots  & \partial^{N - 1}_x f_N \cr
                                      }
\equiv [0, \cdots, N - 1], \label{wronsolneweqnewrep}
\end{equation}
where the symbol $j$ in $[\cdots]$ denote the $j$th derivative of the 
column vector $^t(f_1, \cdots f_N)$.  Then the derivatives of $\tau_N$  are 
described as 

\begin{eqnarray}
& &\tau_{N,x} = [0, \cdots, N - 2, N], 
\label{taunx} \\
& &\tau_{N,z} = \mp [0, \cdots, N - 3, N - 1, N] \pm [0, \cdots, N - 2, 
N + 1], \label{taunz} \\
& &\tau_{N,t} = \pm \frac{1}{2} [0, \cdots, N - 5, N - 3, N- 2, N - 1, N] 
\nonumber \\
& &~~~~~~~~\mp \frac{1}{2} [0, \cdots, N - 4, N - 2, N- 1, N + 1] 
\nonumber \\
& &~~~~~~~~\pm \frac{1}{2} [0, \cdots, N - 3, N - 1, N + 2] 
\nonumber \\
& &~~~~~~~~\mp \frac{1}{2} [0, \cdots, N - 2, N + 3], 
\label{taunt} \\
& & \hspace{3cm} \vdots \hspace{3cm} . \nonumber
\end{eqnarray}
Hence equation(\ref{trilinearneweq}) becomes 

\begin{eqnarray}
\fl & &\pm 8 [0, \cdots, N - 1] \biggl([0, \cdots, N - 1] [0, \cdots, 
N - 3, N, N + 3] \nonumber \\
\fl & &\hspace{2.5cm} - [0, \cdots, N - 2, N] [0, \cdots, N - 3, N - 1, 
N + 3] \nonumber \\
\fl & &\hspace{2.5cm} + [0, \cdots, N - 2, N + 3] [0, \cdots, N - 3, 
N - 1, N]\biggr) \nonumber \\
\fl & &\mp 8 [0, \cdots, N - 1] \biggl([0, \cdots, N - 1] [0, \cdots, 
N - 5, N - 3, N - 2, N, N + 1] \nonumber \\
\fl & &\hspace{2.5cm} - [0, \cdots, N - 2, N] [0, \cdots, N - 5, N - 3, 
N - 2, N - 1, N + 1] \nonumber \\
\fl & &\hspace{2.5cm} + [0, \cdots, N - 2, N + 1] [0, \cdots, N - 5, 
N - 3, N - 2, N - 1, N]\biggr) \nonumber \\
\fl & &\mp 8 [0, \cdots, N - 2, N] \biggl([0, \cdots, N - 1] [0, \cdots, 
N - 3, N, N + 2] \nonumber \\
\fl & &\hspace{3cm} - [0, \cdots, N - 2, N] [0, \cdots, N - 3, N - 1, 
N + 2] \nonumber \\
\fl & &\hspace{3cm} + [0, \cdots, N - 2, N + 2] [0, \cdots, N - 3, 
N - 1, N]\biggr) \nonumber \\
\fl & &\pm 8 [0, \cdots, N - 2, N] \biggl([0, \cdots, N - 1] [0, \cdots, 
N - 4, N - 2, N, N + 1] \nonumber \\
\fl & &\hspace{3cm} - [0, \cdots, N - 2, N] [0, \cdots, N - 4, N - 2, 
N - 1, N + 1] \nonumber \\
\fl & &\hspace{3cm} + [0, \cdots, N - 2, N + 1] [0, \cdots, N - 4, N - 2, 
N - 1, N]\biggr) \nonumber \\
\fl & &\pm 8 \bigl\{[0, \cdots, N - 3, N - 1, N] - [0, \cdots, N - 2, 
N + 1]\bigr\} \nonumber \\
\fl & &\hspace{2.5cm} \times \biggl([0, \cdots, N - 1] [0, \cdots, N - 3, N, 
N + 1] \nonumber \\
\fl & &\hspace{2.8cm} - [0, \cdots, N - 2, N] [0, \cdots, N - 3, N - 1, 
N + 1] \nonumber \\
\fl & &\hspace{2.8cm} + [0, \cdots, N - 2, N + 1] [0, \cdots, N - 3, N - 1, 
N]\biggr) = 0 \label{trilinearneweqnewrep} 
\end{eqnarray}
Since $\biggl(~~~\biggr)$ parts of equation(\ref{trilinearneweqnewrep}) 
are nothing but the Pl\"ucker relations\cite{satosato}, each term of the 
left handside of equation(\ref{trilinearneweqnewrep}) become zero.  This 
completes the proof.  

Thus equation(\ref{neweq}) is proved to be an completely integrable 
system and must have an infinite series of invariants.  In this 
connection we proceed to discuss the Miura transformation of equation 
(\ref{neweq}).  The Miura transformation in the dependent variable of 
equation(\ref{neweq}) is 

\begin{equation}
\phi_x = \psi_x^2 + \sigma \psi_{xx} + \sigma \psi_z
\label{miuratf}
\end{equation}
with $\sigma = \pm i$.  This transformation is the same as the potential 
KP equation(p-KP)'s\cite{kono}.  Then we obtain the modified equation 

\begin{equation}
\fl \psi_{xt} + \frac{1}{4} \psi_{xxxz} + \sigma \psi_{xz} \psi_z + 
\biggl(\frac{1}{2} \psi_x + \frac{1}{2} \psi_{xx} \partial_x^{-1} - 
\frac{1}{4} \sigma \partial_x^{-1} \partial_z\biggr) 
\biggl((\psi_x^2)_z + \sigma \psi_{zz}\biggr) = 0. \label{mneweq}
\end{equation}
We have the Lax pair of equation(\ref{mneweq}) as follows,

\begin{eqnarray}
\fl & &L = \partial_x^2 + 2 \sigma \psi_x \partial_x + \partial_z,
\label{mneweqofL} \\
\fl & &T = \partial_x^2 \partial_z + \sigma \psi_z \partial_x^2 + 
2 \sigma \psi_x \partial_x \partial_z + \frac{1}{2} \partial_z^2 + 
\biggl(- 2 \psi_x \psi_z + \frac{3}{2} \sigma \psi_{xz} + \frac{1}{2} 
\partial_x^{-1} (\psi_x^2)_z \nonumber \\
\fl & &~~~~~- \frac{1}{2} \sigma \partial_x^{-1} 
\psi_{zz}\biggr) \partial_x + \partial_t, \label{mneweqofT}
\end{eqnarray}
where we have replaced $\sigma \to - \sigma$ in equation(\ref{mneweq}).  

Here concluding remarks are in order.  We have obtained the exact 
$N$-soliton solution(\ref{nssofneweq}) and $N \times N$ Wronskian 
solution(\ref{wronsolneweq}) of equation(\ref{neweq}) and have 
constructed the modified equation(\ref{mneweq}) using the Miura 
transformation(\ref{miuratf}).  Moreover, we have constructed the Lax 
pair of equation(\ref{mneweq}).  

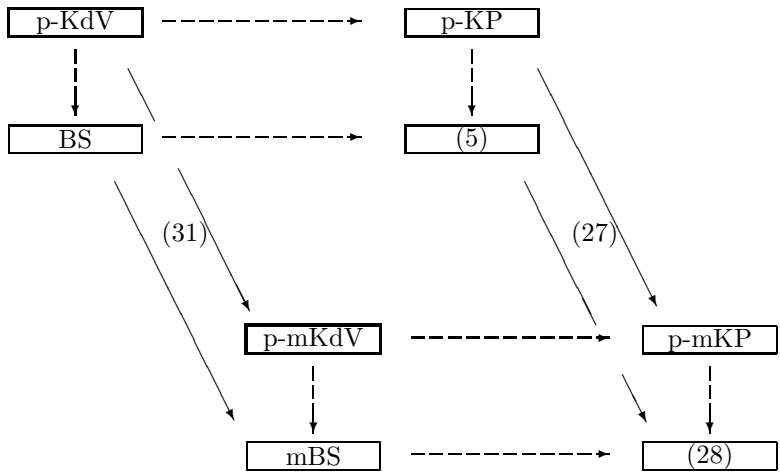
\begin{figure}[h]
\begin{center}
\begin{picture}(100,180)
\put(-100,170){\thicklines\framebox(50,10){p-KdV}}
\put(-42,176){\line(2,0){5}}
\put(-35,176){\line(2,0){5}}
\put(-28,176){\line(2,0){5}}
\put(-21,176){\line(2,0){5}}
\put(-14,176){\line(2,0){5}}
\put(-7,176){\line(2,0){5}}
\put(0,176){\line(2,0){5}}
\put(7,176){\line(2,0){5}}
\put(14,176){\line(2,0){5}}
\put(21,176){\line(2,0){5}}
\put(28,176){\vector(2,0){5}}
\put(50,170){\framebox(50,10){p-KP}}
\put(-75,165){\line(0,-1){5}}
\put(-75,158){\line(0,-1){5}}
\put(-75,151){\vector(0,-1){11}}
\put(75,165){\line(0,-1){5}}
\put(75,158){\line(0,-1){5}}
\put(75,151){\vector(0,-1){11}}
\put(-100,126){\framebox(50,10){BS}}
\put(-42,132){\line(2,0){5}}
\put(-35,132){\line(2,0){5}}
\put(-28,132){\line(2,0){5}}
\put(-21,132){\line(2,0){5}}
\put(-14,132){\line(2,0){5}}
\put(-7,132){\line(2,0){5}}
\put(0,132){\line(2,0){5}}
\put(7,132){\line(2,0){5}}
\put(14,132){\line(2,0){5}}
\put(21,132){\line(2,0){5}}
\put(28,132){\vector(2,0){5}}
\put(50,126){\framebox(50,10){(\ref{neweq})}}
\put(-10,50){\thicklines\framebox(50,10){p-mKdV}}
\put(52,56){\line(2,0){5}}
\put(59,56){\line(2,0){5}}
\put(66,56){\line(2,0){5}}
\put(73,56){\line(2,0){5}}
\put(80,56){\line(2,0){5}}
\put(87,56){\line(2,0){5}}
\put(94,56){\line(2,0){5}}
\put(101,56){\line(2,0){5}}
\put(108,56){\line(2,0){5}}
\put(115,56){\line(2,0){5}}
\put(122,56){\vector(2,0){5}}
\put(140,50){\framebox(50,10){p-mKP}}
\put(15,45){\line(0,-1){5}}
\put(15,38){\line(0,-1){5}}
\put(15,31){\vector(0,-1){11}}
\put(165,45){\line(0,-1){5}}
\put(165,38){\line(0,-1){5}}
\put(165,31){\vector(0,-1){11}}
\put(-10,6){\framebox(50,10){mBS}}
\put(52,12){\line(2,0){5}}
\put(59,12){\line(2,0){5}}
\put(66,12){\line(2,0){5}}
\put(73,12){\line(2,0){5}}
\put(80,12){\line(2,0){5}}
\put(87,12){\line(2,0){5}}
\put(94,12){\line(2,0){5}}
\put(101,12){\line(2,0){5}}
\put(108,12){\line(2,0){5}}
\put(115,12){\line(2,0){5}}
\put(122,12){\vector(2,0){5}}
\put(140,6){\framebox(50,10){(\ref{mneweq})}}
\put(-55,158){\line(1,-2){10}}
\put(-36,120){\vector(1,-2){27}}
\put(-60,115){\vector(1,-2){45}}
\put(100,158){\vector(1,-2){45}}
\put(95,115){\line(1,-2){27}}
\put(132,42){\vector(1,-2){9}}
\put(109,90){\makebox(25,10){(\ref{miuratf})}}
\put(-46,90){\makebox(25,10){(\ref{miuratfofkdvandbs})}}
\end{picture}
\end{center}
\caption{Scheme of extensions of the KdV equation and the mKdV equation.  
There are two directional routes of extensions(dashed arrows): One leads 
us to the p-KP equation and the p-mKP equation.  Another does us to the 
BS equation and the mBS equation.  Equations (5) and (28) are given by 
unifying two routes of extensions.  Four equations in the bottom part of 
this figure are induced by the Miura transformations(solid arrows).}
\label{extensions}
\end{figure}

In the previous paper\cite{fty1} we discussed the construction method 
for the higher dimensional integrable equations.  We obtained the higher 
dimensional equations(the p-KP equation, the BS equation and 
equation(\ref{neweq})) from the p-KdV equation by this method, which is 
schematically depicted in the top part of Figure.\ref{extensions}.  The 
Miura transformation of the p-KdV equation\cite{miura} and the BS 
equation\cite{fsty1,b2} is 

\begin{equation}
\phi_x = \psi_x^2 + \sigma \psi_{xx}.
\label{miuratfofkdvandbs}
\end{equation}
The corresponding modified equations are called the potential modified 
KdV(p-mKdV) equation and the modified BS(mBS) equation, respectively.  The 
latter is 

\begin{equation}
\psi_{xt} + \frac{1}{4} \psi_{xxxz} + \psi_x^2 \psi_{xz} + \frac{1}{2} 
\partial_x^{-1} (\psi_x^2)_z = 0.
\label{mbseq}
\end{equation}
The Miura transformation of the p-KP equation is same as 
equation(\ref{neweq})'s, i.e. equation(\ref{miuratf})\cite{kono}.  In this 
letter, we have given the modified equation(\ref{mneweq}) of 
equation(\ref{neweq}) from the Miura transformation(\ref{miuratf})(see 
Figure.\ref{extensions}).  B. G. Konopelchenko and V. G. Dubrovsky gave 
the Lax pairs of the mKdV equation and the mKP equation\cite{kd}.  
However, the Lax pairs of the mBS equation(\ref{mbseq}) and of 
equation(\ref{mneweq}) were not discussed.  Using the construction method 
for higher dimensional equation, We have obtained the Lax pair of the 
mBS equation(\ref{mbseq})

\begin{eqnarray}
\fl & &L = \partial_x^2 + 2 \sigma \psi_x \partial_x \label{mbseqofL} \\ 
\fl & &T = \partial_x^2 \partial_z + \sigma \psi_z \partial_x^2 + 
2 \sigma \psi_x \partial_x \partial_z + \biggl(- 2 \psi_x \psi_z + 
\frac{3}{2} \sigma \psi_{xz} + \frac{1}{2} \partial_x^{-1} 
(\psi_x^2)_z\biggr) \partial_x + \partial_t \label{mbseqofT}
\end{eqnarray}
and the Lax pair(\ref{mneweqofL}), (\ref{mneweqofT}) of the modified
equation(\ref{mneweq})(see the bottom part of Figure.\ref{extensions}).  
Thus we have succeeded to construct the Lax pairs of the higher 
dimensional modified equations(the m-BS equation and 
equation(\ref{mneweq})), composing the analogous ring of modified systems.  

\section*{References}
\begin{thebibliography}{99}
\bibitem{kd} B. G. Konopelchenko and V. G. Dubrovsky 1983 {\it Phys. Lett.} 
\bibitem{b} O. I. Bogoyavlenskii 1990 {\it Math. USSR. Izv.} {\bf 34} 245
\bibitem{fty1} Yu. S, K. Toda and T. Fukuyama  solv-int/9802005
\bibitem{wtc} J. Weiss, M. J. Tabor and G. Carnevale 1983 {\it J. Math. 
Phys.} {\bf 24} 522
\bibitem{grh} B. Grammaticos, A. Ramani and J. Hietarinta 1994 {\it Phys. 
Lett. A} {\bf 190} 65
\bibitem{hgr} J. Hietarinta, B. Grammaticos and A. Ramani  solv-int/9411003
\bibitem{satosato} M. Sato and Y. Sato 1983 {\it Nonlinear Partial 
Differential Equations in Applied Science}, ed. H. Fujita, P. D. Lax and G. 
Strang (North-Holland) P~259
\bibitem{kono} B. G. Konopelchenko 1982 {\it Phys. Lett.} {\bf 92} {\it A} 
323
\bibitem{miura} R. Miura 1968 {\it J. Math. Phys.} {\bf 9} 1202
\bibitem{fsty1} Yu. S, K. Toda, N. Sasa and T. Fukuyama 1998 
{\it J. Phys. A} accepted (solv-int/9801003) 
\bibitem{b2} O. I. Bogoyavlenskii 1991 {\it Math. USSR. Izv} {\bf 36} 129
{\bf 102} {\it A} 15
\end{thebibliography}
\end{document}